\documentclass[aps,prl,superscriptaddress,showpacs,floatfix,twocolumn]{revtex4}

\usepackage{graphicx}
\usepackage{hyperref}
\usepackage{amsmath}
\usepackage{url}
\usepackage{colordvi}
\usepackage{color}
\usepackage{subfigure}
\usepackage{cancel}
\usepackage{times}

\begin{document}

	
\title{Lepton angular distribution of $W$ boson productions}

\author{Yang Lyu}
\affiliation{Department of Physics, University of Illinois at
Urbana-Champaign, Urbana, Illinois 61801, USA}
\affiliation{Department of Physics, University of California at Berkeley,
Berkeley, California 94720, USA}

\author{Wen-Chen Chang}
\affiliation{Institute of Physics, Academia Sinica, Taipei 11529, Taiwan}

\author{Randall Evan McClellan}
\affiliation{Department of Physics, University of Illinois at
Urbana-Champaign, Urbana, Illinois 61801, USA}
\affiliation{Department of Physical Sciences, Pensacola State College,
Pensacola, Florida 32504, USA}

\author{Jen-Chieh Peng}
\affiliation{Department of Physics, University of Illinois at 
Urbana-Champaign, Urbana, Illinois 61801, USA}

\author{Oleg Teryaev}
\affiliation{Bogoliubov Laboratory of Theoretical Physics,
JINR, 141980 Dubna, Russia}

\date{\today}

\begin{abstract}
The lepton angular distribution coefficients $A_i$ for $Z$ boson
production in $pp$ and $\bar p p$ collisions have been measured at the
LHC and the Tevatron. A recent study showed that many features of the
measured angular distribution coefficients, including the transverse
momentum ($q_T$) and rapidity dependencies and the violation of the
Lam-Tung relation, can be well described using an intuitive geometric
approach.  In this paper, we extend this geometric approach to
describe the angular distribution coefficients for $W$ boson produced
in $\bar{p} p$ collisions at the Tevatron.  We first compare the data
with a perturbative QCD calculation at $\mathcal{O}(\alpha_s^2)$. We
then show that the data and QCD calculations can be well described
with the geometric approach. Implications for future studies at the
LHC energy are also discussed.

\end{abstract}

\pacs{12.38.Lg,14.20.Dh,14.65.Bt,13.60.Hb}

\maketitle

\section{Introduction}
	
Dilepton production in hadron-hadron collision has been studied 
extensively following the pioneering experiment performed
in 1970~\cite{christenson70}.
The mechanism for dilepton production involves a quark annihilating 
with an antiquark, forming a 
vector boson ($\gamma^*, W,$ or $Z$),
which subsequently decays into a pair of leptons.
In the original Drell-Yan model~\cite{drell70},
the vector boson was predicted to be transversely polarized, leading to an
azimuthally symmetric $1+\cos^2\theta$ lepton angular distribution
with respect to the beam axis.
This prediction agreed well with early fixed-target dilepton production data 
where the transverse momentum ($q_T$) of the dilepton is 
low~\cite{kenyon82}. 

The azimuthal symmetry for the lepton angular distribution no longer 
holds for nonzero value of $q_T$, and 
a general expression
for the lepton angular distribution is given as~\cite{lam78}
\begin{equation}\label{eq:qcd_considered}
\frac{d \sigma}{d \Omega} \propto 1 + \lambda \cos^2\theta + \mu 
\sin2\theta\cos\phi + \frac{\nu}{2}\sin^2\theta\cos2\phi
\end{equation}
where $\theta$ and $\phi$ are the lepton polar and azimuthal angles in
the dilepton rest frame. In the original Drell-Yan model
\cite{drell70}, $\lambda = 1$ and $\mu = \nu = 0$. However, the
intrinsic transverse momenta of partons and QCD effects can result in
nonzero values for $\nu$ and $\mu$, while $\lambda$ can also deviate
from unity. It was pointed out by Lam and Tung~\cite{lam78} that the
amount of deviation of $\lambda$ from 1 is twice the value of $\nu$,
namely, $1-\lambda = 2 \nu$. This so-called Lam-Tung relation was
shown to be insensitive to corrections from leading-order QCD
processes~\cite{lam78}.

The Lam-Tung relation was found to be significantly violated in pion-induced 
Drell-Yan experiments \cite{faciano86,guanziroli88,conway89,heinrich91}. 
Many theoretical models~\cite{brandenburg94,eskola94,boer99} were 
proposed to explain this violation. 
In particular, Boer~\cite{boer99} suggested that a novel 
transverse-momentum dependent
(TMD) parton distribution, the Boer-Mulders function~\cite{boer98}, 
can give rise to
a $\cos 2\phi$ azimuthal angular modulation, resulting in a violation
of the Lam-Tung relation. This not only explained the observed violation
of the Lam-Tung relation but also allowed the first extraction of the 
Boer-Mulders functions from pion and proton induced 
Drell-Yan data~\cite{boer99,zhu07}.

Recent high-statistics 
measurements of the lepton angular distribution coefficients in 
$Z$ boson production over a broad range of $q_T$ 
in $pp$ collision at the LHC by the CMS~\cite{cms} and 
ATLAS~\cite{atlas} experiments
revealed a clear violation of the Lam-Tung relation. Since 
TMD effects are only relevant at the low $q_T$ region, the results
from LHC showed that sources other than the Boer-Mulders functions
are responsible for the violations of
the Lam-Tung relation at high $q_T$. Indeed, the fixed-order 
QCD calculations can account for the LHC data rather 
well~\cite{Z_QCD}.

In Ref.~\cite{peng16}, the lepton angular distribution 
in $Z$ boson production was described using
an intuitive geometric approach. Both 
the violation of the Lam-Tung relation and the observed $q_T$ 
dependence of $\lambda$ and $\nu$ could be well described by 
this approach. 
A subsequent paper~\cite{chang17} showed that this approach 
could explain both the $q_T$ and the rapidity dependencies of 
the lepton angular distributions. Several recent papers have also
addressed various aspects of lepton angular distributions in $Z$ boson
production~\cite{Ralston18,oleg19,peng19,peng19b} and the
Drell-Yan process~\cite{werner16,chang19}. 
	
In addition to the lepton angular distribution data for $Z$ boson
production in $\bar p p$~\cite{CDF_Z} and $p p$~\cite{cms,atlas} 
collisions, there are also 
$W$ boson production data in $\bar p p$ collision reported by the
CDF Collaboration~\cite{CDF_W}.
The important roles of the lepton angular distribution in
understanding the mechanisms for $W$ and $Z$ boson production at the
$\bar p p$ collisions at the Tevatron were pointed out in 
Refs.~\cite{mirkes92,mirkes94}.
Unlike the $Z$ boson production where both $l^-$ and $l^+$ decay
products are detected, only the charged lepton from $W$ boson decay
is measured. Consequently, different experimental uncertainties
are encountered in the measurements of lepton angular distributions
in $W$ versus $Z$ boson production. Another important difference 
is that $W$ and
$Z$ boson productions involve different parity-violating couplings. 
Therefore, it is instructive to compare the
lepton angular distribution of $W$ production with that of $Z$
production. In this paper, we extend our previous geometric 
approach of interpreting 
the lepton angular distribution for $Z$ boson production to
$W$ boson production.  
     
This paper is organized as follows. In Sec. II, we briefly describe
the geometric approach and present the implications of this approach
on the lepton angular distribution coefficients of $W$ production.
In Sec. III we compare the CDF data on the angular coefficients 
of $W$ production
with a perturbative QCD calculation. We then show in Sec. IV that
the geometric approach can provide qualitative agreement with the
QCD calculation and the CDF data. We also discuss possible future measurements
at LHC on the angular coefficients of $W$ production. 
We conclude in Sec. V. 

\section{Geometric approach for lepton angular distribution coefficients}
	
In hadron-hadron collision, the angular distribution of charged leptons 
in the $W^{\pm}$ rest frame is 
given by the CDF Collaboration~\cite{CDF_W} as
\begin{equation}
\begin{aligned}[b]\label{eq:full}
\frac{d \sigma}{d \Omega} & \propto (1 + \cos^2\theta) + 
\frac{A_0}{2}(1-3\cos^2\theta)+A_1\sin2\theta\cos\phi \\
&+ \frac{A_2}{2}\sin^2\theta\cos2\phi + A_3\sin\theta\cos\phi 
+ A_4\cos\theta \\
&+ A_5 \sin^2\theta\sin2\phi + A_6\sin2\theta\sin\phi  \\
& + A_7\sin\theta\sin\phi
\end{aligned}
\end{equation}
where $\theta$ and $\phi$ are the polar and azimuthal angles of 
charged lepton in the rest frame of $W$. 
Comparing Eq.~(2) with Eq.~(\ref{eq:qcd_considered}), we obtain
\begin{equation}
\lambda = \frac{2-3A_0}{2+A_0} ,~~~ \mu = \frac{2A_1}{2 + A_0} ,~~~ \nu = 
\frac{2A_2}{2+A_0}
\end{equation}
and the Lam-Tung relation, $1 - \lambda = 2 \nu$, becomes $A_0 = A_2$. 
	
To shed some light on the meaning of the angular distribution coefficients
$A_i$ in Eq. (2), we define three different planes in the rest frame 
of the $W$ boson, as shown in Fig. 1. These planes are (1) the hadron plane formed by the two colliding hadrons' momenta
$\vec{p}_B$ and $\vec{p}_T$.
For the Collins-Soper (C-S) frame~\cite{collins77}, the $\hat{z}$ and 
$\hat{x}$ axes lie in the hadron plane where $\hat{z}$ bisects, with angle
$\beta$, the 
two hadron momentum vectors, $\vec p_B$ and $- \vec p_T$. 
(2) The quark plane formed by $\hat z$ and  
the axis $\hat z^\prime$, along which a pair of quark and antiquark
collide collinearly to produce a $W$ boson at rest. The polar and 
azimuthal angles of $\hat{z}'$ are defined 
as $\theta_1$ and $\phi_1$, respectively in the C-S frame. (3) The lepton plane
defined by the momentum vector of the charged lepton ($l$) and the 
$\hat{z}$ axis. It is worth noting that the definitions of these three
planes and angles are completely general and independent of the 
specific reaction mechanism for producing the $W$ boson.
\begin{figure}[tb]
\includegraphics*[width=\linewidth]{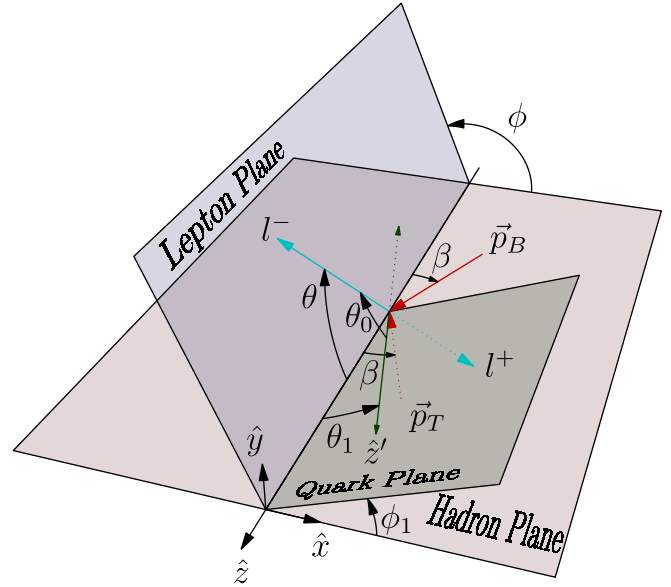}
\caption{Definition of the C-S frame and various 
planes in the rest frame of $W$ boson. The hadron plane
is formed by $\vec P_B$ and $\vec P_T$, the momentum vectors of
the colliding hadrons $B$ and $T$. The $\hat x$ and $\hat z$ axes
of the C-S frame both lie in the hadron plane with
$\hat z$ axis bisecting the angle between 
$\vec P_B$ and $- \vec P_T$ vectors with angle $\beta$.
The quark ($q$) and antiquark ($\bar q$) collide head-on
with equal momenta to form the $W$ boson at rest, while the quark momentum
unit vector $\hat z^\prime$ and the $\hat z$ axis form the quark plane.
The polar and azimuthal angles of $\hat z^\prime$ in the C-S
frame are $\theta_1$ and $\phi_1$. The $l$ and $\nu$ leptons are emitted
back-to-back from $W$ with $\theta$ and $\phi$ specifying the polar and
azimuthal angles of the charged lepton $l$.}
\label{fig:frame}
\end{figure}
	
While Eq. (2) can be derived using the technique of contracting the
leptonic and hadronic tensors~\cite{collins77,mirkes94}, it is instructive to
derive Eq. (2) using a different approach~\cite{peng16}.
In the rest frame of $W$, the charged lepton angular distribution has a very
simple form when it is expressed with respect to the $\hat z^\prime$ axis,
namely, 
\begin{equation}\label{eq:azimuthal}
\frac{d \sigma}{d \Omega}  \propto 1 + a\cos\theta_0 + \cos^2\theta_0~,
\end{equation}
where $\theta_0$ is the polar angle of $l$ with respect to the quark 
momentum as shown in Fig. 1. The forward-backward 
asymmetry parameter, $a$, originates 
from the parity-violating coupling to the $W$ boson. 
For the Drell-Yan process involving the electromagnetic coupling to
a virtual photon, parity is conserved and $a=0$. For $Z$ boson production,
$a=2 A_f A_{f^\prime}$, where $A_f$ is a function of the vector $C^f_V$
and axial-vector $C^f_A$ couplings for $Z$ boson to fermion $f$, as 
discussed in Ref.~\cite{chang17}.
As Eq. (2)
is expressed in terms of the angles $\theta$ and $\phi$, one
can substitute the following trigonometric relation 
\begin{equation}
\cos\theta_0 = \cos\theta\cos\theta_1 + 
\sin\theta\sin\theta_1\cos(\phi-\phi_1)
\end{equation}
into Eq.~(\ref{eq:azimuthal}) to obtain
\begin{equation}
\begin{aligned}[b]\label{eq:derived_full}
\frac{d \sigma}{d \Omega} & \propto (1 + \cos^2\theta) + 
\frac{\sin^2 \theta_1}{2}(1-3\cos^2\theta)\\
&+(\frac{1}{2}\sin2\theta_1 \cos\phi_1)\sin2\theta\cos\phi \\
&+ (\frac{1}{2}\sin^2\theta_1\cos2\phi_1)\sin^2\theta\cos2\phi \\
&+ (a\sin\theta_1\cos\phi_1)\sin\theta\cos\phi + (a\cos \theta_1)\cos\theta \\
&+ (\frac{1}{2}\sin^2\theta_1\sin 2\phi_1) \sin^2\theta\sin2\phi \\
&+ (\frac{1}{2}\sin2\theta_1\sin \phi_1)\sin2\theta\sin\phi  \\
& + (a\sin\theta_1 \sin\phi_1)\sin\theta\sin\phi,
\end{aligned}
\end{equation}
which contains all angular terms in Eq. (2). Comparing Eq. (2) with 
Eq. (6), we find that all lepton angular distribution 
coefficients, $A_i$, can
be expressed in terms of $\theta_1$, $\phi_1$, and $a$ as follows:
\begin{align}
A_0 &= \langle \sin^2\theta_1 \rangle & A_1 &= \langle \frac{1}{2} \sin
2\theta_1 \cos \phi_1 \rangle \nonumber \\
A_2 &=  \langle \sin^2\theta_1 \cos 2\phi_1 \rangle & A_3 &= \langle a
\sin \theta_1 \cos \phi_1 \rangle \nonumber \\
A_4 &=  \langle a \cos \theta_1 \rangle & A_5 &=  \langle \frac{1}{2}
\sin^2\theta_1 \sin 2\phi_1 \rangle \nonumber \\
A_6 &= \langle \frac{1}{2} \sin 2\theta_1 \sin\phi_1 \rangle &
A_7 &=  \langle a \sin\theta_1 \sin\phi_1 \rangle.
\label{eq7}
\end{align}
The brackets indicate that the measured coefficients are obtained by 
averaging over all events. In this way the lepton angular 
distribution coefficients could be related to the 
polar and azimuthal angles, $\theta_1$ and $\phi_1$,  
of the quark axis in the $W$ rest frame.

One difference between $W$ and $Z$ boson productions is that $W$ boson 
production
maximally violates parity. 
The $V-A$ coupling for the $W$ boson implies a
$(1+\cos \theta_0)^2$ or $(1-\cos \theta_0)^2$ lepton angular distribution
for $W^+$ and $W^-$ productions in Eq. (4). Hence $|a| =2$, and Eq. (7) 
implies that
the ranges of the angular distribution coefficients are
\begin{equation}
\begin{aligned}[b]
0\leq A_0 \leq 1 \ &,\  ~~-1\leq A_2 \leq 1 \\
-2\leq A_3 \leq 2 \ &,\  ~~-2\leq A_4 \leq 2,
\end{aligned}
\label{eq8}
\end{equation}
\noindent where we only consider $A_0, A_2, A_3, A_4$, which were measured
by the CDF Collaboration. 
Equation (7) also shows that $A_0 \geq A_2$. Therefore, 
when the Lam-Tung relation, $A_0 = A_2$, is violated, 
$A_2$ can only be smaller, not greater, than $A_0$~\cite{peng19}.
	
We emphasize that the expressions of Eqs. (6)-(8) are completely
general, independent of the choice of the reference frame. The exact values
of the polar angles $(\theta, \theta_1)$ and azimuthal angles $(\phi, \phi_1)$
do depend, in general, on the choice of the reference frame. There are many 
different choices for the $W$ boson rest frame in the literature, 
including the Collins-Soper frame~\cite{collins77}, the 
Gottfried-Jackson frame~\cite{gottfried64}, the
U-channel frame~\cite{lam78}, the helicity frame, and the Mustraal 
frame~\cite{was17}, corresponding
to different choices for the orientations of the axes.

It is possible to find the values of $\theta_1$ and $\phi_1$ for
certain specific $W$ boson production processes~\cite{peng16,chang17}.
Consider first a $\mathcal{O}(\alpha_s)$ process, $q \bar q \to
WG$, in which a quark from one hadron annihilates with an antiquark
from another hadron to form a $W$ boson. A hard gluon $G$ is emitted
from either the quark or the antiquark, resulting in a nonzero
transverse momentum for the $W$.  It is easy to see that in the C-S
frame, $\theta_1$ must be identical to the angle $\beta$ in
Fig. 1~\cite{chang17}.  Emission of a gluon from one of the colliding
partons cannot change the momentum of the other parton, which
continues to move along the $\vec p_B$ or $\vec p_T$ direction. Hence,
the $q \bar q$ collision axis ($\hat z^\prime$ in Fig. 1) is along the
$\vec p_B$ or $\vec p_T$ direction, making an angle $\beta$ with
respect to the $\hat z$ axis in the C-S frame. It is straightforward
to obtain~\cite{chang17}
\begin{equation}
\sin^2\theta_1 = \sin^2\beta = q_T^2/(Q^2 + q_T^2),
\label{eq9}
\end{equation}
\noindent where $q_T$ and $Q$ are the transverse momentum and mass of the
$W$, respectively. Since the quark plane and the hadron plane both contain
$\hat z$ and $\vec p_B$ (or $\vec p_T$), these two planes coincide and 
$\phi_1$ must vanish for this process.
  
For the $qG \to q^\prime W$ Compton process, the value of $\theta_1$ was
found~\cite{chang17,thews} to be given approximately as
\begin{equation}
\sin^2\theta_1 = 5q_T^2/(Q^2 + 5q_T^2),
\label{eq10}
\end{equation}
\noindent while $\phi_1$ remains zero. It is interesting to note that
at $\mathcal{O}(\alpha_s)$, Eq. (7) shows that the Lam-Tung
relation, $A_0 = A_2$, is satisfied since $\phi_1 = 0$.  At $\mathcal{O}(\alpha_s^2)$ or higher, the quark plane is in general different from
the hadron plane due to the emission of more than one
jet~\cite{chang17,peng19b}. Hence $\phi_1 \ne 0$ and the Lam-Tung
relation will be violated.

\section{Comparison between the CDF data and perturbative QCD calculation}

The CDF Collaboration reported the measurement of the $A_2$ and $A_3$ angular
coefficients of the $W$ boson production in $\bar p p$ collision at 
$\sqrt s = 1.8$ TeV~\cite{CDF_W}. From the detection of the charged lepton
momentum from the $W \to e \nu$ and $W \to \mu \nu$ decays and the missing
transverse energy 
$\cancel{E}_{T}$, the azimuthal angle $\phi$ of the charged 
lepton in the C-S frame is measured. However, the polar angle $\theta$ of
the charged lepton cannot be uniquely determined due to a twofold ambiguity
resulting from the unknown longitudinal momentum of the neutrino. The $W$ 
boson events satisfy the requirements
\begin{eqnarray}
E^e_T (P^\mu_T) &\geq&  20~\mbox{GeV}~~~~~~~~~~\cancel{E}_{T} > 20~\mbox{GeV} 
\nonumber \\
|\eta^{e,\mu}| &\leq& 1~~~~~~~~~~~~15 <q^W_T < 105~\mbox{GeV}~,
\label{eq:CDFcuts}
\end{eqnarray}
where $\eta$ is the pseudorapidity of the charged lepton.

\begin{figure}[tb]
\includegraphics*[width=\linewidth]{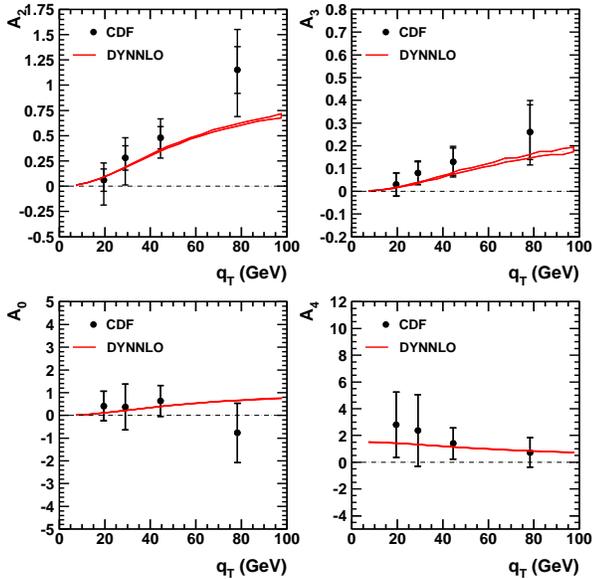}
\caption{Comparison between the CDF $W$ boson angular coefficient 
data~\cite{CDF_W} with $\mathcal{O}(\alpha_s^2)$ QCD calculation. $q_T$ is the transverse momentum of the $W$ boson.
The coefficients $A_2$ and $A_3$ display
both the statistical uncertainties (inner error bars) and
the total uncertainties (outer error bars). For $A_0$ and $A_4$, only statistical error bars are provided by CDF. The $\mathcal{O}(\alpha_s^2)$ QCD calculation, shown as the solid curves, utilized the
DYNNLO code and the CT14NNLO PDFs
for proton and antiproton.}
\label{fig_CDF}
\end{figure}

From the measurement of the azimuthal angle
of the charged lepton in the C-S frame from the $W \to e \nu$ and
$W \to \mu \nu$ decays, the angular coefficients $A_2$ and $A_3$ were 
extracted. With much reduced sensitivity, the coefficients $A_2$ and $A_4$
were also measured. Figure 2 shows the CDF data on $A_0, A_2,
A_3$ and $A_4$ versus the transverse momentum $q_T$ of the $W$ boson.
Both the statistical and total (statistical plus systematic) uncertainties
are shown for $A_2$ and $A_3$. As the statistical uncertainties for
$A_0$ and $A_4$ are large, no estimates for their systematic uncertainties were
provided by the CDF Collaboration. 

We first compare the CDF results with perturbative QCD calculation at $\mathcal{O}(\alpha_s^2)$. For this calculation, we utilize the 
DYNNLO code~\cite{DYNNLO} version 1.5~\cite{v1_5}, which provides the
differential cross sections for the Drell-Yan process and $W/Z$ boson 
production. The CT14NNLO parton distribution functions were used for the
proton and antiproton in this calculation. From the calculated
$d\sigma/d\Omega$ differential angular distribution, the $A_i$
angular coefficients can be evaluated by taking the appropriate
moments, namely~\cite{mirkes94},
\begin{align}
A_0 &=  4 - 10 \langle \cos^2\theta \rangle & A_2 &=  10 \langle
\sin^2\theta \cos 2\phi \rangle \nonumber \\
A_3 &=  4 \langle \sin \theta \cos \phi \rangle &  A_4 &=  4 \langle
\cos \theta \rangle~,
\label{eq:eq11} 
\end{align}
where $\langle f(\theta,\phi) \rangle$ denotes the moment of 
$f(\theta,\phi)$, i.e., 
\begin{equation}
\langle f(\theta,\phi) \rangle = \frac{\int f(\theta,\phi) 
\frac{d\sigma}{d\Omega}
d\Omega}{\int \frac{d\sigma}{d\Omega} d\Omega}~.
\end{equation}
Equation (12) is obtained by using the orthogonality property of the
various angular distribution terms in Eq. (2).
The results of the calculation for $A_0, A_2, A_3, A_4$ are shown as solid curves in 
Fig.~\ref{fig_CDF}. The finite vertical widths of the curves reflect the 
variations when using other two PDF sets, NNPDF31nnlo and MMHT2014nnlo,
for the calculation.

We note some qualitative features of the QCD calculation. 
As $q_T \to 0$, Eqs.~(\ref{eq9}) and~(\ref{eq10}) show
that $\theta_1 =0$. Equation~(\ref{eq7}) requires that all $A_i$ 
except $A_4$ vanish when $\theta_1 = 0$. This is confirmed
by the QCD calculation shown in Fig.~\ref{fig_CDF}. Moreover, the upper
and lower bounds listed in Eq.~(\ref{eq8}) are satisfied by the QCD
calculation. 
The agreement between the calculation and the CDF
data is quite good. We note that the present results are also 
in good agreement with an earlier QCD calculation performed by members of
the CDF Collaboration~\cite{Errede}.  

\begin{figure}[tb]
\includegraphics*[width=\linewidth]{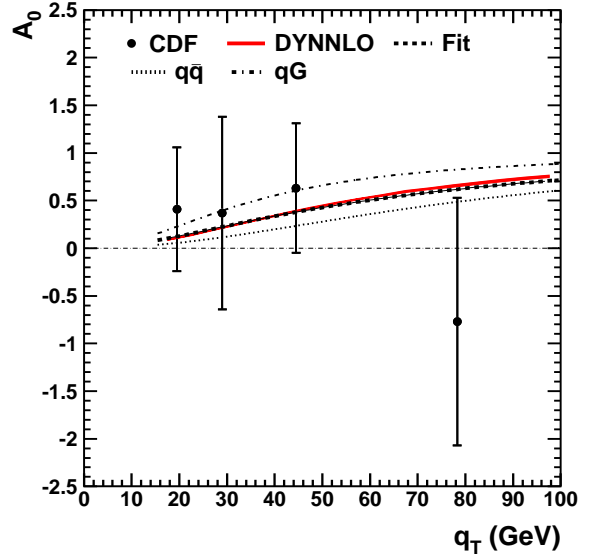}
\caption{Comparison between the $\mathcal{O}(\alpha_s^2)$ QCD 
(solid curve) and the
geometric approach (dashed curve) for the calculations
of $A_0$. The dotted and
dot-dashed curves correspond to the contribution
from the $q \bar q$ and $q G$ subprocess, respectively,
in the geometric approach. The CDF data points~\cite{CDF_W} are also 
displayed.}
\label{fig_A0}
\end{figure}

\section{Interpretation of the angular coefficients with the geometric
approach}

\begin{figure}[tb]
\includegraphics*[width=\linewidth]{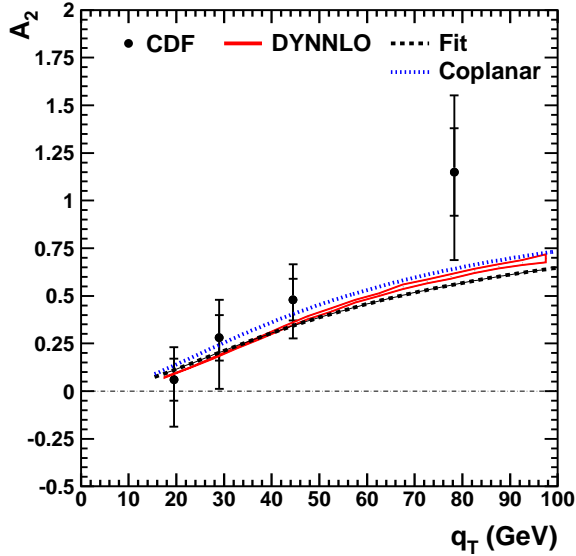}
\caption{Comparison between the $\mathcal{O}(\alpha_s^2)$
QCD (solid curve) and the
geometric approach (dashed curve) for the calculations
of $A_2$. The dotted 
curve corresponds to the calculation of the geometric
model when the non-coplanarity angle $\phi_1$ is set at zero.
The CDF data points~\cite{CDF_W} are also displayed.}
\label{fig_A2}
\end{figure}

While the $\mathcal{O}(\alpha_s^2)$ QCD calculation can describe
the angular coefficients of $W$ production very well as shown in
Fig.~2, it is instructive to examine how well the intuitive geometric
approach discussed in Sec. II can reproduce the main features of
the data. In the earlier studies of the $Z$ boson
production~\cite{peng16,chang17}, the high statistics of the LHC data
made it possible to use the data to constrain some parameters in the
geometric approach. Unfortunately, the large uncertainty for the $W$
boson production data from CDF greatly limits the sensitivity of using
the data to test the geometric approach. Therefore, we use instead the
$\mathcal{O}(\alpha_s^2)$ QCD results to check whether the
geometric approach can adequately describe the angular coefficients
for $W$ boson production.

We start with the $A_0$ angular coefficient. Equation (7) shows that $A_0$
is given by the values of $\sin^2\theta_1$ averaged over the different 
processes. At $\mathcal{O}(\alpha_s)$, Eqs. (9) and (10) give the $q_T$
dependence of $\sin^2\theta_1$ for the quark-antiquark annihilation 
and the quark-gluon Compton process, respectively. 
The dotted and dot-dashed curves in Fig. 3 correspond to Eqs. (9) and
(10), respectively. Note that the $qG$ process alone overestimates $A_0$, while $q\bar{q}$ underestimates it.
As the $q \bar q$ annihilation and the $q G$ Compton
processes involve different initial states, they contribute incoherently 
to the $W$ production. The dashed curve in Fig. 3 is obtained with
the following expression
\begin{equation}\label{eq:A0}
A_0 = f \frac{q_T^2}{Q^2+q_T^2} + (1-f)\frac{5q_T^2}{Q^2+5q_T^2},
\end{equation}
where $f$ represents the fraction of $q\bar{q}$ process, and $1-f$ is 
the fraction of the $qG$ process. The best fit to the QCD calculation
gives $f=0.610 \pm 0.002$, which is consistent with the expectation that
the $q \bar q$ annihilation process dominates the $q G$ process in $p \bar p$
collision. The excellent agreement between the geometric approach and the
QCD calculation suggests that Eqs. (9) and (10) are capable of reproducing
the main features of the QCD calculation for $A_0$.

We next consider the $A_2$ angular coefficient. If Lam-Tung relation 
is satisfied, then $A_0 = A_2$. Figure 4 compares the $\mathcal{O}(\alpha_s^2)$ QCD calculation for $A_2$ (solid curve) with the result of $A_0$ from the
geometric approach (dotted curve) obtained with Eq. (14). While the 
agreement is reasonable, the QCD calculation is consistently below 
the dotted curve. This indicates that the Lam-Tung relation, $A_0 = A_2$,
is violated. In the geometric approach, this implies that the angle
$\phi_1$ is nonzero, which leads to a smaller $A_2$ than $A_0$, as 
shown in Eq. (7). A nonzero $\phi_1$ implies that
the quark and hadron planes are not coplanar. This non-coplanarity
is caused by higher-order QCD processes at $\mathcal{O}(\alpha_s^2)$ or higher, in
which multiple partons accompany the $W$ boson in the final state.
To account for the nonzero $\phi_1$ angle, we
use the following expression: 
\begin{equation}\label{eq:A2}
A_2 =  \Big( f \frac{q_T^2}{Q^2+q_T^2} + (1-f) 
\frac{5q_T^2}{Q^2+5q_T^2}\Big) \cos 2\phi_1
\end{equation}
The dashed curve in Fig. 4 corresponds to Eq. (15) with $f = 0.610$
obtained from the $A_0$ data discussed above, and the best-fit value
for $\cos 2\phi_1 = 0.905 \pm 0.004$. This corresponds to an average
non-coplanarity angle, $\phi_1$, of $12.6^\circ$. The improved agreement
between the geometric model and QCD calculation using this nonzero
$\phi_1$ angle indicates the effects of $\mathcal{O}(\alpha_s^2)$ or higher,
which allows the hadron plane to deviate from the quark plane. An analogous
situation was found for $Z$ boson production and discussed in~\cite{chang17}.
The good agreement between the simple calculation using Eq. (15) and the
sophisticated QCD calculation again illustrates the useful insight
provided by the geometric approach for understanding the angular coefficient
in $W$ boson production.

\begin{figure}[tb]
\includegraphics*[width=\linewidth]{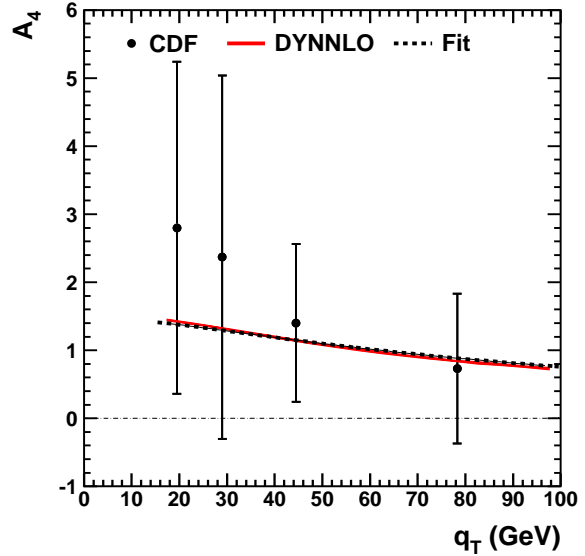}
\caption{Comparison between the $\mathcal{O}(\alpha_s^2)$ QCD (solid curve) and the
geometric model (dashed curve) for $A_4$.
The CDF data points~\cite{CDF_W} are also displayed.}
\label{fig_A4}
\end{figure}

We consider next the parity-violating angular coefficient $A_4$. From Eq. (7),
the expression for $A_4$ is the product of the forward-backward asymmetry
parameter $a=\pm 2$ and $\cos \theta_1$. Since the expressions for 
$\sin^2 \theta_1$ are given by Eqs. (9) and (10) for the $q \bar q$ and
$q G$ processes, we use the following expression for $A_4$
\begin{equation}
A_4 = 2 r_4 \Big( f\frac{Q}{(Q^2+q_T^2)^{\frac{1}{2}}}  + 
(1-f)\frac{Q}{(Q^2+5q_T^2)^{\frac{1}{2}}}  \Big)
\end{equation}
where the factor of 2 on the right-hand side signifies the magnitude of the
forward-backward asymmetry, $|a| = 2$. The parameter $r_4$, which has a 
magnitude less than 1, is to account for the fact that the sign of $a$ is
either positive or negative, depending on whether it is 
$q \bar q \to W$ or $\bar q q \to W$ process, as discussed in 
Ref.~\cite{chang17} for the analogous $Z$ boson production. 
Depending on the relative weight of these two contributions, governed by the
partonic distributions of quarks and antiquarks in the proton and antiproton,
the magnitude of $A_4$ is expected to be reduced from the partial 
cancellation effect. The parameter $r_4$ accounts for such a partial 
cancellation effect. The dashed curve in Fig. 5 shows that Eq. (16),
using the best-fit value of $r_4 = 0.738 \pm 0.002$,  is
in excellent agreement with the QCD calculation. The large uncertainty of
the $A_4$ measurement from CDF prevents a conclusive comparison between the
QCD and geometric model calculation with the data. In fact, the constraint
$A_4 < 2$ from Eq. (8), marginally violated by the central values of the 
data points at the lowest two $q_T$ values, is satisfied by the QCD 
calculation. The fact that the simple calculation of Eq. (16) can describe
the $\mathcal{O}(\alpha_s^2)$ QCD calculation very well 
again indicates the adequacy of the simple
geometric model in understanding the main features of the angular
coefficients in $W$ boson production.

\begin{figure}[tb]
\includegraphics*[width=\linewidth]{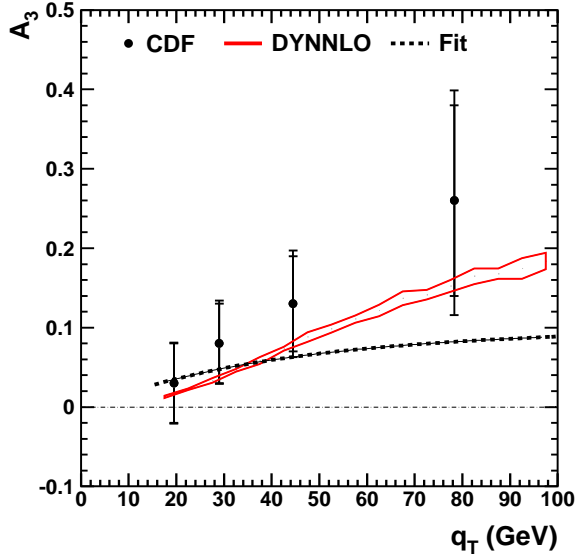}
\caption{Comparison between the $\mathcal{O}(\alpha_s^2)$ QCD 
(solid curve) and the
geometric model (dashed curve) for $A_3$.
The CDF data points~\cite{CDF_W} are also displayed.}
\label{fig_A3}
\end{figure}

We turn to the $A_3$ coefficient next. As shown in Eq. (7), $A_3$ involves
all three quantities, $\theta_1, \phi_1,$ and $a$. The partial cancellation
effects discussed for $A_4$ are also expected for $A_3$. From Eqs. (7), (9),
and (10), we use the following expression for $A_4$ in the geometric model:
\begin{equation}
A_3 =  2 r_3\Big( f\frac{q_T}{(Q^2+q_T^2)^{\frac{1}{2}}} + 
(1-f) \frac{\sqrt{5}q_T}{(Q^2+5q_T^2)^{\frac{1}{2}}}  \Big)\cos \phi_1
\end{equation}
where the factor of 2 is again the forward-backward asymmetry parameter 
for $W$ boson production, and $\phi_1 = 12.6^\circ$ was obtained in the 
previous analysis of $A_2$. Since $A_3$ is an odd function under the 
$\phi_1 \leftrightarrow \pi - \phi_1$ exchange, a large cancellation effect
is expected \cite{chang17}. Therefore, we expect the value of the 
reduction factor, $r_3$, to be small.  The dashed curve in Fig.~\ref{fig_A3} is the best fit
to the $\mathcal{O}(\alpha_s)$ QCD calculation using Eq. (17). 
Indeed, the value of $r_3$
is found to be quite small, $r_3 = 0.0540 \pm 0.001$, confirming a very large
cancellation effect. Moreover, the agreement
between the calculations of the geometric model and the 
$\mathcal{O}(\alpha_s^2)$ QCD is not very
good. This suggests that the simple assumption that $r_4$ is independent
of $q_T$ is no longer a good assumption in the presence of strong
cancellation effects. Nevertheless, the general feature that $A_3$ increases
with $q_T$ can still be described by the geometric model.

We conclude this section by discussing the prospect for collecting and
analyzing $W$ boson angular distribution data at the LHC. The high luminosity
and high center-of-mass energy at the LHC allow a precise measurement
of $W$ and $Z$ boson production. Indeed, the lepton angular distribution
data for $Z$ production reported by CMS and ATLAS have demonstrated a much
higher precision and broader $q_T$ coverage than at Tevatron. It is still
challenging to measure $W$ boson angular distribution due to the missing
neutrinos. As a result, only the $W$ polarization parameters, $f_L, f_R$ 
and $f_0$ in the helicity
frame have been measured~\cite{CMS_W,ATLAS_W} so far. 
Nevertheless, as shown by the CDF Collaboration, at least the
$A_2$ and $A_3$ coefficients in the C-S frame could be measured with 
adequate precision
even at Tevatron. It is anticipated that LHC could at least allow a very
precise measurement of $A_2$ and $A_3$ coefficients for $W$ production.
As discussed in a recent paper~\cite{peng19b}, the $A_2$ 
coefficient is expected to have
very different $q_T$ distributions for $Z$ boson plus single jet or multiple
jets. Similar expectation also holds for $W$ plus jets production at the LHC.
In particular, the $A_2$ values for $W$ plus multiple-jets events are expected
to be smaller than for the $W$ plus single-jet events. This is due to the
nonzero values of $\phi_1$ for a multiple-jets events, while a single-jet
$W$ production event must have $\phi_1 = 0$. Equation (7) then implies that
$A_2$ for multiple-jet events must be smaller than that for single-jet events.
This prediction remains to be tested by the LHC experiments.

We note that a recent paper suggests the possibility of reconstructing
the $W$ decay angular distribution in the Mustraal frame~\cite{was17}.
From an analysis of the Monte Carlo data at the LHC energy, it was shown
that the Mustraal frame has the interesting property that all angular
coefficients except $A_4$ have vanishing values. An inspection of Eq. (4)
shows that the $z$ axis of the Mustraal frame coincides with the 
$\hat z^\prime$ axis. Thus far, all existing data on $W$ and $Z$ boson
production from LHC are analyzed in either the C-S frame or the helicity
frame. Future analysis of these data in the Mustraal frame would be of
considerable interest.

\section{Summary and Conclusions}
    
In this paper, we have extended the geometric approach to describe 
the angular distribution coefficients of $W$ boson production at the CDF.
In this geometric approach, first discussed in~\cite{peng16},
all of the eight lepton angular distribution coefficients can be expressed  
as trigonometric expressions involving three quantities: $a$, $\theta_1$, 
and $\phi_1$. The quantity $\theta_1$ refers to the polar angle of the
collinear quark-antiquark axis in the $W$ boson rest frame, $\phi_1$, the
non-coplanarity angle between the plane formed by the two hadrons and the
lepton plane containing the leptons from the $W$ decay. The parity-violating parameter $a$ has a magnitude of 2 for $W$ production. These
trigonometric expressions lead to a set of upper and lower bounds for the
various angular coefficients, as well as some relationships between these
angular coefficients. In particular, the Lam-Tung relation refers to the
equality of the $A_0$ and $A_2$ coefficients when the non-coplanarity angle
$\phi_1$ vanishes. The violation of the Lam-Tung relation is then
attributed to a nonzero $\phi_1$ angle, resulting in $A_0 > A_2$. For
the $q \bar q$ annihilation and $qG$ Compton processes at $\mathcal{O}(\alpha_s)$, $\phi_1$ vanishes, and the Lam-Tung relation is valid. For processes
at $\mathcal{O}(\alpha_s^2)$ or higher, $\phi_1$ can be nonzero and the Lam-Tung
relation will be violated.

We first compare the CDF angular coefficient data with 
$\mathcal{O}(\alpha_s^2)$ QCD calculation. Although the
statistical precision of the CDF data is only marginal, the general
features of the data are in good agreement with the QCD calculations. 
We then compare the QCD results with the expressions obtained
from the geometric model in order to determine several parameters in this
model. Good agreement between the QCD calculations and
the geometric approach is obtained. We also confirm that the QCD
calculations as well as the geometric approach satisfy the upper and
lower bounds derived for the angular coefficients, and the
inequality $A_0 > A_2$. The implication of this study for $W$ production
at LHC is also discussed. In particular, a high precision
measurement of $A_2$ for $W$ plus jets events is feasible and of much interest.

We emphasize that this geometric approach is developed to provide 
some simple intuitive insights for understanding the angular distribution
coefficients for $W$ and $Z$ boson production. It is certainly not
a substitute for the rigorous perturbative QCD calculations. 
The good agreement between the geometric approach and the perturbative
QCD calculation as well as the data is reassuring that the geometric model
has some merits in understanding the main features of the lepton angular
distributions, including their transverse momentum dependence and the
violation of the Lam-Tung relation, in an intuitive fashion.
We expect that
this geometric approach can also be extended to other hard processes,
including the Drell-Yan process,
quarkonium production, $e^- e^+$ collision, and deep-inelastic 
scattering.

\section*{Acknowledgment}

This research
was supported in part by the U.S. National Science Foundation 
Grant No. PHYS18-22502 and
the Ministry of Science and Technology of Taiwan.


\begin{thebibliography}{90}
\bibitem{christenson70} J. H. Christenson {\em et al.}, Phys. Rev.
Lett. {\bf 25}, 1523 (1970).

\bibitem{drell70} S. D. Drell and T. M. Yan, Phys. Rev. Lett. {\bf 25}, 316
   (1970); Ann. Phys. (NY) {\bf 66}, 578 (1971).

\bibitem{kenyon82} I. R. Kenyon, Rep. Prog. Phys. {\bf 45}, 1261 (1982).

\bibitem{lam78} C. S. Lam and W. K. Tung, Phys. Rev. {\bf D 18}, 2447
(1978).

\bibitem{faciano86} S. Falciano {\em et al.}, Z. Phys. C {\bf 31},
513 (1986).

\bibitem{guanziroli88} M. Guanziroli {\em et al.}, Z. Phys. C {\bf 37}, 
545 (1988).

\bibitem{conway89} J. S. Conway {\em et al.}, Phys. Rev.
{\bf D 39}, 92 (1989).

\bibitem{heinrich91} J. G. Heinrich {\em et al.}, Phys. Rev.
{\bf D 44}, 1909 (1991).

\bibitem{brandenburg94} A. Brandenburg, S. J. Brodsky, V. V. Khoze, and D. M\"{u}ller, Phys. Rev. Lett. {\bf 73}, 939 (1994).

\bibitem{eskola94} K. Eskola, P. Hoyer, M. V\"{a}ntinnen, and R. Vogt, Phys. Lett. B {\bf 333}, 526 (1994).

\bibitem{boer99} D. Boer, Phys. Rev. D {\bf 60}, 014012 (1999).

\bibitem{boer98} D. Boer and P. J. Mulders, Phys. Rev. D {\bf 57}, 
5780 (1998).

\bibitem{zhu07} L. Y. Zhu {\em et al.}, Phys. Rev. Lett. {\bf 99}, 082301 
(2007); {\bf 102}, 182001 (2009).

\bibitem{cms} V. Khachatryan {\em et al.} (CMS Collaboration), Phys.
Lett. {\bf B 749}, 187 (2015).

\bibitem{atlas} G. Aad {\em et al.} (ATLAS Collaboration), J. High Energy
Phys. {\bf 08} (2016) 159.

\bibitem{Z_QCD} R. Gauld {\em et al.}, J. High Energy Phys. {\bf 11} (2017) 003.

\bibitem{peng16} J. C. Peng, W. C. Chang, R. E. McClellan, and O. Teryaev,
Phys. Lett. {\bf B 758}, 384 (2016).

\bibitem{chang17} W. C. Chang, R. E. McClellan, J. C. Peng, and O. Teryaev,
Phys. Rev. D {\bf 96}, 054020 (2017).

\bibitem{Ralston18} J. C. Martens, J. P. Ralston, and J. P. Tapia Takaki,
Eur. Phys. J. C {\bf 78}, 5 (2018).

\bibitem{oleg19} M. Gavrilova and O. Teryaev,
Phys. Rev. D {\bf 99}, 076013 (2019).

\bibitem{peng19} J. C. Peng, D. Boer, W. C. Chang,
R. E. McClellan, and O. Teryaev,
Phys. Lett. B {\bf 789}, 356 (2019).

\bibitem{peng19b} J. C. Peng, W. C. Chang,
R. E. McClellan, and O. Teryaev,
Phys. Lett. B {\bf 797}, 134895 (2019).

\bibitem{werner16} M. Lambertsen and W. Vogelsang,
Phys. Rev. D {\bf 93}, 114013 (2016).

\bibitem{chang19} W. C. Chang, R. E. McClellan, J. C. Peng, and O. Teryaev,
Phys. Rev. D {\bf 99}, 014032 (2019).

\bibitem{CDF_Z} T. Aaltonen {\em et al.} (CDF Collaboration), Phys. Rev.
Lett. {\bf 106}, 241801 (2011).

\bibitem{CDF_W} D. Costa {\em et al.} (CDF Collaboration), Phys. Rev. D
{\bf 73}, 052002 (2006).

\bibitem{mirkes92} E. Mirkes, Nucl. Phys. {\bf B387}, 3 (1992).

\bibitem{mirkes94} E. Mirkes and J. Ohnemus, Phys. Rev. D {\bf 50},
5692 (1994).
 
\bibitem{collins77} J. C. Collins and D. E. Soper, Phys. Rev. D {\bf 16},
    2219 (1977).

\bibitem{gottfried64} K. Gottfried and J. D. Jackson, Nuovo Cimento
{\bf 33}, 309 (1964).

\bibitem{was17} E. Richter-Was and Z. Was, Eur. Phys. J. C {\bf 77},
111 (2017).

\bibitem{thews} R. L. Thews, Phys. Rev. Lett. {\bf 43}, 987 (1979).

\bibitem{DYNNLO} S. Catani and M. Grazzini, Phys. Rev. Lett.
{\bf 98}, 222002 (2007); S. Catani {\em et al.}, Phys. Rev. Lett.
{\bf 103}, 082001 (2009).

\bibitem{v1_5} DYNNLO v1.5, http://theory.fi.infn.it/ grazzini/dy.html. 

\bibitem{Errede} J. Strologas and S. Errede,
Phys. Rev. D {\bf 73}, 052001 (2006).

\bibitem{CMS_W} S. Chatrchyan {\em et al.} (CMS Collaboration),
Phys. Rev. Lett. {\bf 107}, 021802 (2011).

\bibitem{ATLAS_W} M. Aaboud {\em et al.} (ATLAS Collaboration),
Eur. Phys. J. C {\bf 79}, 535
(2019).


\end{thebibliography}
\end{document}